# Exclusive Semileptonic Decays of $B_c$ Meson

*Rohit Dhir and R.C. Verma*

Department of Physics, Punjabi University,
Patiala-147002, INDIA.

**Abstract**

We present a detailed analysis of exclusive semileptonic decays of $B_c$ meson involving pseudoscalar and vector mesons in the BSW framework, by investigating the effects of the flavor dependence on the average transverse quark momentum inside a meson. Branching ratios of exclusive semileptonic decays of $B_c$ meson are predicted.

PACS number(s): 13.20.-v, 13.20.He, 12.39.Ki

## 1. Introduction

The discovery of the $B_c$ meson by the collider detector at Fermilab (CDF) [1] opens up some interesting investigations concerning the structure of strong and weak interactions. The properties of the $B_c$ meson are of special interest [2], since it is the only heavy meson consisting of two heavy quarks with different flavors. This difference of quarks flavor forbids annihilation in to gluons. A peculiarity of the $B_c$ decays, with respect to the decays of $B$ and $B_s$ mesons, is that both the quarks may involve in its weak decays. There are quite a few numbers of theoretical works studying various leptonic, semileptonic and hadronic decay channels of $B_c$ mesons in different models [3-14]. Their estimates of $B_c$ decay rates indicate that the *c*-quark give dominant contribution as compared to *b*-quark decays. From experimental point of view, study of weak decays of $B_c$ meson is quite important for the determination of Cabibbo–Kobayashi–Maskawa (CKM) elements. More detailed information about its decay properties are expected in the near future at LHC and other experiments.

In our recent work [14], we have investigated the effects of flavor dependence on $B_c \rightarrow P/V$ form factors, caused by possible variation of average transverse quark momentum ($\omega$) in a meson, here $P$ and $V$ stands for pseudoscalar and vector meson respectively. Employing the Bauer, Stech, and Wirbel (BSW) framework [15], we then predicted the branching ratios of $B_c$ meson decays involving pseudoscalar and vector mesons. In the present paper, we extend our analysis to investigate such effects on exclusive semileptonic decays of $B_c$ meson. We observe that the branching ratios of semileptonic $B_c$ decays get enhanced when such flavor dependent effects are included.

The present paper is organized as follows: In Section 2, we describe the factorization hypothesis and kinematic formula. Section 3 deals with $B_c$ form factors in the BSW model. We study the effects of flavor dependence of $\omega$ on $B_c \rightarrow P/V$ form factors in Section 4. Finally the branching ratios of exclusive semileptonic decays of $B_c$ meson are predicted. Section 5 contains summary and conclusions.

## 2. Factorization scheme

In the standard factorization scheme, the weak current $J_\mu$ is given by

$$J_\mu = (\bar{u}\ \bar{c}\ \bar{t}\ )\, \gamma_\mu(1\text{-}\gamma_5) \begin{pmatrix} d' \\ s' \\ b' \end{pmatrix}, \tag{1}$$

and $d', s', b'$ are mixture of the *d, s* and *b* quarks, as given by the Cabibbo-Kobayashi-Maskawa (CKM) matrix [16]. Matrix elements of the currents are defined [15] as,

$$\langle V | J_\mu | B_c \rangle = \frac{2}{m_{B_c} + m_V} \varepsilon_{\mu\nu\rho\sigma} \varepsilon^{*\nu} P_{B_c}^{\rho} P_V^{\sigma} V(q^2)$$
$$+ i\{\varepsilon_\mu^* (m_{B_c} + m_V) A_1(q^2) - \frac{\varepsilon^* \cdot q}{m_{B_c} + m_V} (P_{B_c} + P_V)_\mu A_2(q^2) \tag{2}$$
$$- \frac{\varepsilon^* \cdot q}{q^2} 2 m_V q_\mu A_3(q^2)\} + i \frac{\varepsilon^* \cdot q}{q^2} 2 m_V q_\mu A_0(q^2),$$
$$\langle P | J_\mu | B_c \rangle = (P_{B_c} + P_P - \frac{m_{B_c}^2 - m_P^2}{q^2} q)_\mu F_1(q^2) + \frac{m_{B_c}^2 - m_P^2}{q^2} q_\mu F_0(q^2), \tag{3}$$

where $\varepsilon_\mu$ denotes the polarization vector of the outgoing vector meson, $q_\mu = (P_{B_c} - P_P)_\mu$, $F_1(0) = F_0(0)$, $A_3(0) = A_0(0)$ and

$$A_3(0) = \frac{m_{B_c} + m_V}{2 m_V} A_1(0) - \frac{m_{B_c} - m_V}{2 m_V} A_2(0). \tag{4}$$

In general, the semileptonic decay amplitude $A(B_c \to X l \bar{\nu})$ can be expressed as

$$A_{SL}(B \to X) = \frac{G_F}{\sqrt{2}} |V_{Qq}|^2 L^\mu H_\mu, \tag{5}$$

where

$$\begin{aligned} L^\mu &= \bar{u}(k_2) \gamma^\mu (1 - \gamma_5) \upsilon(k_1), \\ H_\mu &= \langle X | J_\mu | B \rangle, \end{aligned} \tag{6}$$

for $X$ = $P$ or $V$ and $k_1 = k_l$, $k_2 = k_\nu$ if the decaying quark is a $c(b)$ quark. $|V_{Qq}|^2$ is the appropriate CKM matrix elements for $Q \to q$ transiton.

The decay widths of $B_c \to P l \bar{\nu}$ and $B_c \to V l \bar{\nu}$ can be expressed as a function of the four momentum transfer ($q^2$) between initial and final hadrons. The semileptonic decay width for $B_c \to P l \bar{\nu}$ is given by the formula

$$\Gamma(B_c \to P l \bar{\nu}) = \frac{G_F^2}{(2\pi^3)} |V_{Qq}|^2 \int_{m_l^2}^{q^2} dq^2 \left( \frac{q^2 - m_l^2}{q^2} \right)^2 \frac{\sqrt{\lambda(m_{B_c}^2, m_P^2, q^2)}}{24 m_{B_c}^3} (m_{B_c}^2 - m_P^2)^2 \times \left[ \frac{3 m_l^2}{2 q^2} F_0^2(q^2) + \left( 1 + \frac{m_l^2}{2 q^2} \right) \frac{\lambda(m_{B_c}^2, m_P^2, q^2)}{(m_{B_c}^2 - m_P^2)^2} F_1^2(q^2) \right] \tag{7}$$

where $m_l$ is the mass of the lepton and $0 \le q^2 \le q^2_{\max} = (m_{B_c} - m_X)^2$ and $\lambda(m^2_{B_c}, m^2_X, q^2) = (m^2_{B_c} + m^2_X - q^2) - 4m^2_{B_c} m^2_X$ is related to the three momentum of the daughter meson in the rest frame of $B_c$ meson by $P_X = \frac{\sqrt{\lambda(m^2_{B_c}, m^2_X, q^2)}}{2m_{B_c}}$.

The total decay width for $B_c \to V l\bar{\nu}$ can be defined as the sum of longitudinal and transverse decay widths given by

$$\Gamma(B_c \to V l\bar{\nu}) = \sum_{i=L,\pm} \Gamma_i(B_c \to V l\bar{\nu}). \tag{8}$$

The longitudinal decay width $(\Gamma_L)$ is defined as

$$\Gamma_L(B_c \to V l\bar{\nu}) = \frac{G_F^2}{(2\pi^3)} |V_{Qq}|^2 \int_{m_l^2}^{q^2} dq^2 \left(\frac{q^2 - m_l^2}{q^2}\right)^2 \frac{\sqrt{\lambda(m^2_{B_c}, m^2_V, q^2)}}{24m^3_{B_c}} \times \left[\frac{3m_l^2}{2q^2} \lambda(m^2_{B_c}, m^2_V, q^2) A_0^2(q^2) + \left(1 + \frac{m_l^2}{2q^2}\right) |H_0|^2\right] \tag{9}$$

and the transeverse decay width $(\Gamma_T = \Gamma_+ + \Gamma_-)$ is defined as

$$\Gamma_\pm(B_c \to V l\bar{\nu}) = \frac{G_F^2}{(2\pi^3)} |V_{Qq}|^2 \int_{m_l^2}^{q^2} dq^2 \left(\frac{q^2 - m_l^2}{q^2}\right)^2 \frac{\sqrt{\lambda(m^2_{B_c}, m^2_V, q^2)}}{24m^3_{B_c}} \left(1 + \frac{m_l^2}{2q^2}\right) |H_\pm|^2. \tag{10}$$

The helicity amplitudes $H_0$ and $H_\pm$ are given by

$$H_0 = \frac{(m_{B_c} + m_V)}{2m_V} \left\{ (m^2_{B_c} - m^2_V - q^2) A_1(q^2) - \frac{\lambda(m^2_{B_c}, m^2_V, q^2)}{(m_{B_c} + m_V)^2} A_2(q^2) \right\}, \tag{11}$$

and

$$H_\pm = \left\{ \frac{\lambda(m^2_{B_c}, m^2_V, q^2)}{(m_{B_c} + m_V)} V(q^2) \mp (m_{B_c} + m_V) A_1(q^2) \right\}. \tag{12}$$

It has been pointed out in the BSW2 model [17] that consistency with the heavy quark symmetry requires certain form factors such as $F_1$, $A_0$, $A_2$ and $V$ to have dipole $q^2$ dependence and $A_1$ to have monopole $q^2$ dependence, i.e.

$$F_0(q^2) = \frac{F_0(0)}{(1 - q^2/m_S^2)}, \quad F_1(q^2) = \frac{F_1(0)}{(1 - q^2/m_V^2)^2}, \quad A_0(q^2) = \frac{A_0(0)}{(1 - q^2/m_P^2)^2},$$

$$A_1(q^2) = \frac{A_1(0)}{(1-q^2/m_A^2)}, \; A_2(q^2) = \frac{A_2(0)}{(1-q^2/m_A^2)^2} \text{ and } V(q^2) = \frac{V(0)}{(1-q^2/m_V^2)^2} \tag{13}$$

with appropriate pole masses $m_i$.

**3. Form factors in BSW model**

We employ the BSW model for evaluating the meson form factors. In this model, the initial and final state mesons are given by the relativistic bound states of a quark $q_1$ and an antiquark $\bar{q}_2$ in the infinite momentum frame [15],

$$\begin{aligned} \left|\mathbf{P}, m, j, j_Z\right\rangle = \sqrt{2}(2\pi)^{3/2} \sum_{S_1 S_2} \int d^3 p_1 d^3 p_2 \delta^3(\mathbf{P} - \mathbf{p_1} - \mathbf{p_2}) \\ \times \psi_m^{j,j_Z}(\mathbf{p_{1T}}, x, s_1, s_2) a_1^{S_1^+}(\mathbf{p_1}) b_2^{S_2^+}(\mathbf{p_2}) \left|0\right\rangle \; , \end{aligned} \tag{14}$$

where $P_\mu = (P_0, 0, 0, P)$ with $P \to \infty$, $x$ denotes the fraction of the longitudinal momentum carried by the non-spectator quark $q_1$, and $\mathbf{p_{1T}}$ denotes its transverse momentum:

$$x = p_{1Z}/P, \; \mathbf{p_{1T}} = (p_{1x}, p_{1y}).$$

Though $B_c \to PV$ decays involve $F_1(q^2)$ and $A_0(q^2)$ only, we calculate all the form factors appearing in the expression (7) and (8) to later investigate their flavor dependence.

By expressing the current $J_\mu$ in terms of the annihilation and creation operators, the form factors are given by the following integrals:

$$F_0^{B_c P}(0) = F_1^{B_c P}(0) = \int d^2 p_T \int_0^1 (\psi_P^*(\mathbf{p_T}, x) \psi_{B_C}(\mathbf{p_T}, x)) dx, \tag{15}$$

$$A_0^{B_c V}(0) = A_3^{B_c V}(0) = \int d^2 \mathbf{p_T} \int_0^1 dx (\psi_V^{*1,0}(\mathbf{p_T}, x) \sigma_Z^{(1)} \psi_{B_c}(\mathbf{p_T}, x)), \tag{16}$$

$$V(0) = \frac{m_{q_1(B_c)} - m_{q_1(V)}}{m_{B_c} - m_V} I, \tag{17}$$

and

$$A_1(0) = \frac{m_{q_1(B_c)} + m_{q_1(V)}}{m_{B_c} + m_V} I, \tag{18}$$

where

$$I = \sqrt{2} \int d^2 \mathbf{p_T} \int_0^1 \frac{dx}{x} (\psi_V^{*1,-1}(\mathbf{p_T}, x) i \sigma_y^{(1)} \psi_{B_c}(\mathbf{p_T}, x)), \tag{19}$$

$m_{q_1(B_c)}$ and $m_{q_1(V)}$ denote masses of the non-spectator quarks participating in the quark decay process.

The Meson wave function is given by

$$\psi_m(\mathbf{p_T}, x) = N_m \sqrt{x(1-x)} \exp(-\mathbf{p_T^2}/2\omega^2) \exp(-\frac{m^2}{2\omega^2}(x - \frac{1}{2} - \frac{m_{q_1}^2 - m_{q_2}^2}{2m^2})^2), \tag{20}$$

where $m$ denotes the meson mass and $m_i$ denotes the $i^{th}$ quark mass, $N_m$ is the normalization factor and $\omega$ is the average transverse quark momentum, $\langle \mathbf{p_T^2} \rangle = \omega^2$.

The form factors are sensitive to the choice of $\omega$, which is treated as a free parameter. In the BSW model [15], the form factors are calculated by taking $\omega$=0.40 GeV for all the mesons and $m_u = m_d = 0.35$ GeV, $m_s = 0.55$ GeV, $m_c = 1.7$ GeV and $m_b = 4.9$ GeV. The $B_c \to P$ form factors thus obtained are given in column 3 of Table I and $B_c \to V$ form factors are given in Table II. Using them, we obtain the branching ratios for various $B_c$ decays, with $\tau_{B_c} = 0.46$ ps as given in column 2 of Table III. We also give the relative transverse and longitudinal decay widths in columns 2 and 3 of Table IV. We make the following observations:

1. Naively, one may expect the bottom conserving modes to be kinematically suppressed. However, the large values of the form factors as well as CKM matrix element for $c \to u, s$ transitions assures the higher branching ratios for this mode as compared to the $b \to u, c$ transitions.

2. On the other hand, the decays involving $b \to u, c$ transitions are suppressed due to the small values of the CKM matrix elements. The small values of the $b \to u, c$ transitions form factors due to less overlap of the initial and final state wave functions for $\omega$ = 0.40 GeV, as shown in Fig. I, further suppresses the bottom changing modes.

3. We find that the bottom conserving and charm changing transitions are dominant: $B(B_c^+ \to B_s^0 e\bar{\nu}_e)$ = 6.04×10$^{-3}$, $B(B_c^+ \to B_s^0 \mu\bar{\nu}_\mu)$ = 5.80×10$^{-3}$, $B(B_c^+ \to B_s^{*0} e\bar{\nu}_e)$ = 1.14%, $B(B_c^+ \to B_s^{*0} \mu\bar{\nu}_\mu)$ = 1.08%, $B(B_c^+ \to B^{*0} e\bar{\nu}_e)$ = 6.55×10$^{-4}$, and $B(B_c^+ \to B^{*0} \mu\bar{\nu}_\mu)$ = 6.28×10$^{-4}$. Among the bottom changing transitions $B(B_c^- \to \eta_c e\bar{\nu}_e)$= 6.11×10$^{-4}$, and $B(B_c^- \to J/\psi e\bar{\nu}_e)$ = 8.72×10$^{-4}$ modes dominate, but are small as compared to the bottom conserving modes.

4. The mass difference between electron and muon has negligible effect on $b \to u, c$ decay modes, while the branching ratios of $c \to u, s$ decay modes changes rougly by 5%. The dominant semileptonic decay modes involving the $\tau$-muon are: $B(B_c^- \to \eta_c \tau \bar{\nu}_\tau)$ = 1.96×10$^{-4}$, and $B(B_c^- \to J/\psi\, \tau \bar{\nu}_\tau)$ = 2.04×10$^{-4}$.

5. We observe that all the semileptonic decays, except bottom changing and charm conserving ($\Delta b$ =1, $\Delta$C= 0, $\Delta S$ = -1) decays, are largely longitudinal in character with 10% to 21% contribution from the transverse nature. Relative longitudinal and transverse decay widths are almost comparable for bottom changing and charm conserving decays.

## 4. Effects of flavor dependence on $B_c \to V/P$ form factors

In the previous work [14], we have investigated the possible flavor dependence in $B_c \to P$ form factors and consequently in $B_c \to PP$ decay widths. We wish to point out that the parameter $\omega$, being dimensional quantity, may show flavor dependence. Therefore, it may not be justified to take the same $\omega$ for all the mesons. Following the analysis described in [14], we estimate $\omega$ for different mesons from $|\psi(0)|^2$ i.e. square of the wave function at origin, using the following relation based on the dimensionality arguments

$$|\psi(0)|^2 \propto \omega^3 . \tag{21}$$

$|\psi(0)|^2$ is obtained from the hyperfine splitting term for the meson masses [18],

$$|\psi(0)|^2 = \frac{9 m_i m_j}{32 \alpha_s \pi}(m_V - m_P), \tag{22}$$

where $m_V$ and $m_P$ respectively denotes masses of vector and pseudoscalar mesons composed of $i$ and $j$ quarks. The meson masses fix quark masses (in GeV) to be $m_u = m_d = 0.31$, $m_s = 0.49$, $m_c = 1.7$, and $m_b = 5.0$ for $\alpha_s(m_b) = 0.19$, $\alpha_s(m_c) = 0.25$, and $\alpha_s = 0.48$ (for light flavors $u$, $d$ and $s$).
Except for $B_c^*$, all the meson masses required are available experimentally. Theoretical estimates for hyperfine splitting $m_{B_c^*} - m_{B_c}$ obtained in different quark models [19, 20] range from 65 to 90 MeV. For the present work, we take $m_{B_c^*} - m_{B_c}$ =73 MeV obtained in [19], which has been quite successful in giving charmonium and bottomonium mass spectra. Calculated numerical values of $|\psi(0)|^2$ are listed in column 2 of Table V. We use the well measured form factor [21] $F_0^{DK}(0) = 0.78 \pm 0.04$ to determine $\omega_D$ = 0.43 GeV which in turn yields $\omega$ for other mesons given in column 3 of Table V. Obtained form factors for $B_c \to P$ transitions are given in column 4 of Table I and for $B_c \to V$ transitions are given in Table VI. We find that all the form factors get significantly enhanced due to the large overlap of $B_c$ and the final state meson as shown in Fig. II.

### 4.1 Numerical branching ratios

Using the flavor dependent form factors we predict the branching ratios and relative decay widths of $B_c \to X\, l\bar{\nu}_l$ decays, which are given in column 3 of Table III and in columns 4 and 5 of Table IV respectively. We observe the following:

1. The branching ratios of both bottom changing ($c \to u,\ s$ transitions) as well as bottom conserving ($b \to u,\ c$ transitions) modes get significantly enhanced. However, the bottom conserving and charm changing mode still remain dominant with $B(B_c^+ \to B_s^0\, e\bar{\nu}_e)$ = 1.54%, $B(B_c^+ \to B_s^0\, \mu\bar{\nu}_\mu)$ = 1.48%, $B(B_c^+ \to B_s^{*0}\, e\bar{\nu}_e)$ = 3.58%, $B(B_c^+ \to B_s^{*0}\, \mu\bar{\nu}_\mu)$ = 3.40%, $B(B_c^+ \to B^{*0}\, e\bar{\nu}_e)$ = 0.203%, and $B(B_c^+ \to B^{*0}\, \mu\bar{\nu}_\mu)$ = 0.195%.

2. Among the bottom changing modes, the dominant branching ratios are $B(B_c^- \to \eta_c e\bar{\nu}_e)$=0.61%, and $B(B_c^- \to J/\psi e\bar{\nu}_e)$ = 1.10%. The branching ratios of semileptonic decay modes involving the $\tau$-muon also get enhanced and the dominating decays are: $B(B_c^- \to \eta_c \tau\bar{\nu}_\tau)$= 0.195%, and $B(B_c^- \to J/\psi\, \tau\bar{\nu}_\tau)$ = 0.26%. For the sake of the comparison, we list branching ratios of semileptonic decays of $B_c$ meson obtained by other models in Table III.

3. We observe that all semileptonic decays, except bottom changing and charm conserving decays, remain largely longitudinal in character with an increased (11% to 30%) contribution from the transverse nature as compared to the relative decay widths at $\omega$=0.40 GeV. In case of bottom changing and charm changing decays ($\Delta b$ =1, $\Delta C$= 0, $\Delta S$ = -1), contributions from relative longitudinal and transverse decay widths are nearly equal.

### 5. Summary and conclusions

In this paper, we have employed the BSW relativistic quark model to study the semileptonic weak decays of $B_c$ meson. We have also investigated the flavor dependence of $\omega$, hence consequently of the form factors and the branching ratios for bottom changing as well as bottom conserving decay modes. We draw the following conclusions.

1. One naively expects the decays involving $c \to u,\ s$ transitions (bottom conserving and charm changing) to be kinematically suppressed, however, the large values of the CKM matrix elements as well as form factors overly compensates the suppression. Due to the less overlap of the initial and the final state wave functions, the form factors involving the bottom changing transitions are small as compared to those of the bottom conserving transitions. As a result the decays involving $b \to u, s$ transitions (bottom changing) get suppressed in comparison to the bottom conserving decays.

2. Initially, branching ratios for both the modes are obtained by taking the usual value of $\omega$ = 0.40 GeV for all the mesons. For bottom conserving and charm changing modes, their form factors yield $B(B_c^+ \to B_s^0\, e\bar{\nu}_e)$ = $6.04\times10^{-3}$, $B(B_c^+ \to B_s^0\, \mu\bar{\nu}_\mu)$ = $5.80\times10^{-3}$, $B(B_c^+ \to B_s^{*0}\, e\bar{\nu}_e)$ = 1.14%, $B(B_c^+ \to B_s^{*0}\, \mu\bar{\nu}_\mu)$ = 1.08%, $B(B_c^+ \to B^{*0}\, e\bar{\nu}_e)$ = $6.55\times10^{-4}$, and $B(B_c^+ \to B^{*0}\, \mu\bar{\nu}_\mu)$ = $6.28\times10^{-4}$. The dominant branching ratios for bottom changing decays are: $B(B_c^- \to \eta_c e\bar{\nu}_e)$= $6.11\times10^{-4}$ and $B(B_c^- \to J/\psi e\bar{\nu}_e)$ = $8.72\times10^{-4}$.

3. We have investigated the effects of possible flavor dependence of $\omega$ on form factors for $B_c$ transition form factors which get significantly enhanced for bottom changing as well as for bottom conserving decays. However, bottom conserving decays remain dominant with higher branching ratios: $B(B_c^+ \to B_s^0\, e\bar{\nu}_e)$ = 1.54%, $B(B_c^+ \to B_s^0\, \mu\bar{\nu}_\mu)$ = 1.48%, $B(B_c^+ \to B_s^{*0}\, e\bar{\nu}_e)$ = 3.58%, $B(B_c^+ \to B_s^{*0}\, \mu\bar{\nu}_\mu)$ = 3.40%, $B(B_c^+ \to B^{*0}\, e\bar{\nu}_e)$ = 0.203%, and $B(B_c^+ \to B^{*0}\, \mu\bar{\nu}_\mu)$ = 0.195%, while branching ratios of bottom changing modes are also increased to $B(B_c^- \to \eta_c e\bar{\nu}_e)$ =0.61% and $B(B_c^- \to J/\psi e\bar{\nu}_e)$ = 1.10%.

4. The branching ratios of semileptonic decay modes involving the $\tau$-muon also get enhanced and the dominating decays are: $B(B_c^- \to \eta_c \tau\bar{\nu}_\tau)$= 0.195%, and $B(B_c^- \to J/\psi\, \tau\bar{\nu}_\tau)$ = 0.26%.

5. We have also investigated the effects of flavor dependence of $\omega$ on the relative transverse and longitudinal decay rates. We find, in comparison to the results at $\omega$ = 0.40 GeV, that the transverse rate gets enhanced and the longitudinal rate decreases for all decays. Except for bottom changing and charm conserving decays, relative longitudinal decay widths remain large as compared to relative transverse decay widths for all decays.

**Figure captions**

**Fig. I.** Overlap of wave functions for $B_c \rightarrow \overline{D}^*$ decays at $\omega_{B_c}$ = $\omega_{D*}$ = 0.40 GeV.

**Fig. II.** Overlap of wave functions for $B_c \rightarrow \overline{D}^*$ decays at $\omega_{B_c}$ = 0.96 GeV and $\omega_{D*}$ = 0.43 GeV.

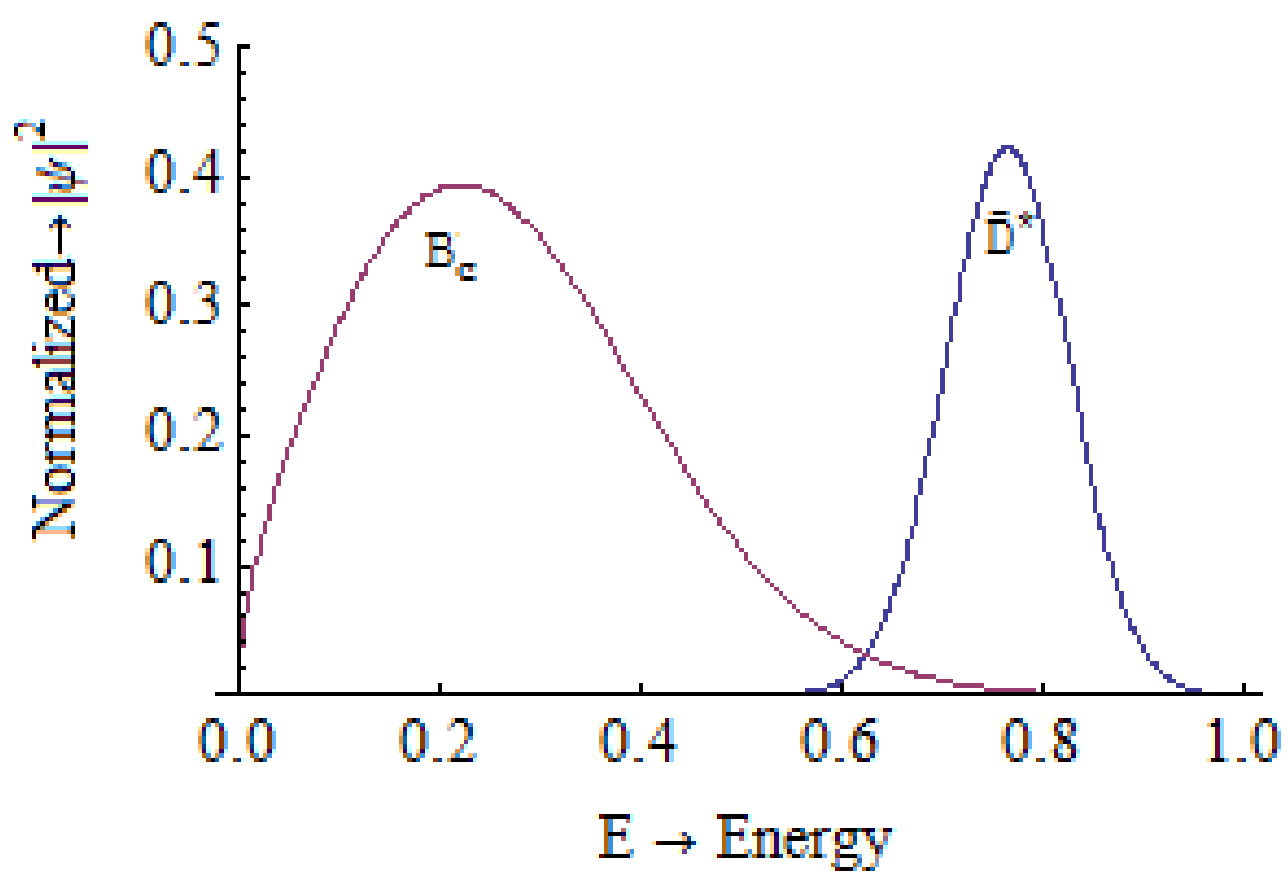

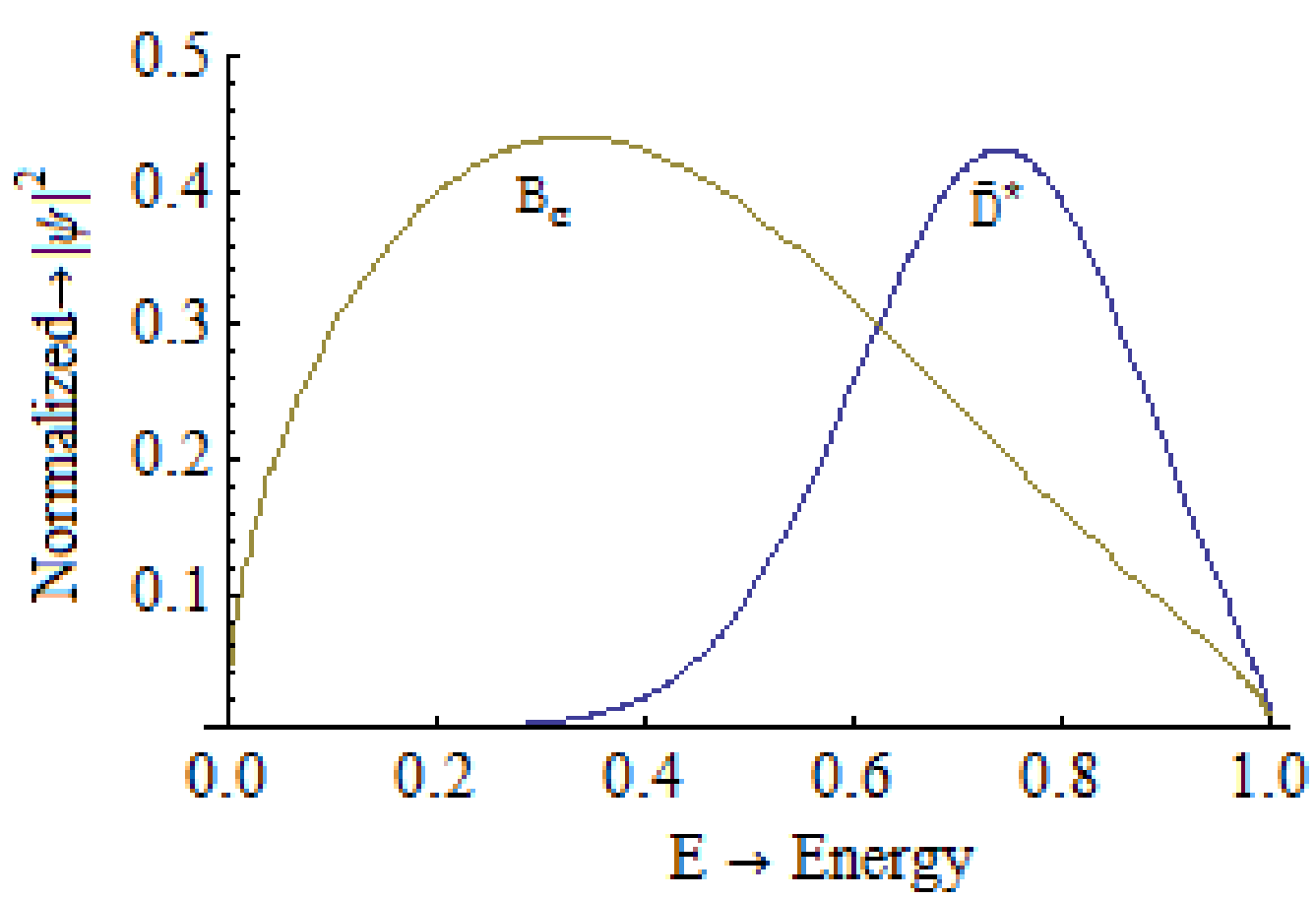

**Tables I. Form factors of $B_c \to P$ transition**

| Modes | Transition | This work | |
|---|---|---|---|
| | | $F_1^{B_cP}(0)$ ($\omega$ = 0.40 GeV) | $F_1^{B_cP}(0)$ (using flavor dependent $\omega$) |
| $\Delta b$ = 0, $\Delta C$ = -1, $\Delta S$ = -1 | $B_c \to B_s$ | 0.35 | 0.55 |
| | $B_c \to B$ | 0.28 | 0.41 |
| $\Delta b$ =1, $\Delta C$ = 0, $\Delta S$ = -1 | $B_c \to D$ | 0.015 | 0.075 |
| | $B_c \to D_s$ | 0.021 | 0.15 |
| $\Delta b$ =1, $\Delta C$ = 1, $\Delta S$ = 0 | $B_c \to \eta_c(c\bar{c})$ | 0.19 | 0.58 |

**Tables II. Form factors of $B_c \to V$ transition ($\omega$ = 0.40 GeV)**

| Modes | Transition | $V(0)$ | $A_0(0)$ | $A_1(0)$ | $A_2(0)$ |
|---|---|---|---|---|---|
| $\Delta b$ = 0, $\Delta C$ = -1, $\Delta S$ = -1 | $B_c \to B_s^*$ | 2.45 | 0.37 | 0.40 | 0.68 |
| | $B_c \to B^*$ | 2.23 | 0.31 | 0.31 | 0.35 |
| $\Delta b$ =1, $\Delta C$ = 0, $\Delta S$ = -1 | $B_c \to D^*$ | 0.025 | 0.016 | 0.015 | 0.013 |
| | $B_c \to D_s^*$ | 0.032 | 0.022 | 0.020 | 0.019 |
| $\Delta b$ =1, $\Delta C$ = 1, $\Delta S$ = 0 | $B_c \to J/\psi(c\bar{c})$ | 0.24 | 0.17 | 0.17 | 0.17 |

**Tables III. Branching ratios (in $\frac{\tau_{B_c}\,(\mathrm{ps})}{0.46}$ %) of $B_c \to X\, l\bar{\nu}_l$ decays**

| Decays | This Work | | [4] | [5] | [6] | [8] | [10] |
|---|---|---|---|---|---|---|---|
| | **At $\omega$ = 0.40 GeV** | **Using flavor dependent $\omega$** | | | | | |
| $\Delta b = 0$, $\Delta C = -1$, $\Delta S = -1$ | | | | | | | |
| $B_c^+ \to B^0\, e\bar{\nu}_e$ | $3.29\times10^{-2}$ | $6.87\times10^{-2}$ | 0.114 | 0.046 | 0.34 | 0.06 | 0.071 |
| $B_c^+ \to B^0 \mu\bar{\nu}_\mu$ | $3.19\times10^{-2}$ | $6.67\times10^{-2}$ | 0.109 | 0.044 | - | - | - |
| $B_c^+ \to B_s^0\, e\bar{\nu}_e$ | 0.604 | 1.54 | 1.64 | 1.06 | 4.03 | 0.8 | 1.10 |
| $B_c^+ \to B_s^0\, \mu\bar{\nu}_\mu$ | 0.580 | 1.48 | 1.55 | 1.02 | - | - | - |
| $B_c^+ \to B^{*0} e\bar{\nu}_e$ | $6.55\times10^{-2}$ | 0.203 | 0.149 | 0.11 | 0.58 | 0.19 | 0.063 |
| $B_c^+ \to B^{*0}\, \mu\bar{\nu}_\mu$ | $6.28\times10^{-2}$ | 0.195 | 0.141 | 0.11 | - | - | - |
| $B_c^+ \to B_s^{*0} e\bar{\nu}_e$ | 1.14 | 3.58 | 2.15 | 2.35 | 5.06 | 2.3 | 2.37 |
| $B_c^+ \to B_s^{*0}\, \mu\bar{\nu}_\mu$ | 1.08 | 3.40 | 2.02 | 2.22 | - | | - |
| $\Delta b = 1$, $\Delta C = 0$, $\Delta S = -1$ | | | | | | | |
| $B_c^+ \to D^0 e\bar{\nu}_e$ | $1.28\times10^{-5}$ | $3.02\times10^{-4}$ | $3.00\times10^{-3}$ | - | $4\times10^{-3}$ | $3.0\times10^{-4}$ | $3.5\times10^{-3}$ |
| $B_c^+ \to D^0 \mu\bar{\nu}_\mu$ | $1.28\times10^{-5}$ | $3.01\times10^{-4}$ | $2.99\times10^{-3}$ | - | - | - | - |
| $B_c^+ \to D^0\, \tau\bar{\nu}_\tau$ | $9.88\times10^{-5}$ | $2.33\times10^{-4}$ | $2.11\times10^{-3}$ | - | $2\times10^{-3}$ | | $2.1\times10^{-3}$ |
| $B_c^+ \to D^{*0} e\bar{\nu}_e$ | $3.64\times10^{-5}$ | $1.2\times10^{-3}$ | $4.58\times10^{-3}$ | - | 0.018 | $8\times10^{-3}$ | $3.8\times10^{-3}$ |
| $B_c^+ \to D^{*0}\, \mu\bar{\nu}_\mu$ | $3.64\times10^{-5}$ | $1.11\times10^{-3}$ | $4.58\times10^{-3}$ | - | - | - | - |
| $B_c^+ \to D^{*0}\, \tau\bar{\nu}_\tau$ | $2.23\times10^{-5}$ | $7.79\times10^{-4}$ | $2.69\times10^{-3}$ | - | $8\times10^{-3}$ | - | $2.2\times10^{-3}$ |
| $\Delta b = 1$, $\Delta C = 1$, $\Delta S = 0$ | | | | | | | |
| $B_c^- \to \eta_c e\bar{\nu}_e$ | $6.11\times10^{-2}$ | 0.608 | 0.686 | 0.48 | 0.75 | 0.15 | 0.81 |
| $B_c^+ \to \eta_c \mu\bar{\nu}_\mu$ | $6.10\times10^{-2}$ | 0.606 | 0.684 | 0.48 | - | - | - |
| $B_c^+ \to \eta_c\, \tau\bar{\nu}_\tau$ | $1.96\times10^{-2}$ | 0.195 | 0.195 | 0.17 | 6.23 | - | 0.22 |
| $B_c^+ \to J/\psi e\bar{\nu}_e$ | $8.72\times10^{-2}$ | 1.10 | 1.52 | 1.54 | 1.9 | 1.47 | 2.07 |
| $B_c^+ \to J/\psi\, \mu\bar{\nu}_\mu$ | $8.68\times10^{-2}$ | 1.09 | 1.51 | 1.54 | - | - | - |
| $B_c^+ \to J/\psi\, \tau\bar{\nu}_\tau$ | $2.04\times10^{-2}$ | 0.264 | 0.38 | 0.41 | 0.48 | - | 0.49 |

## Tables IV. Relative decay widths of $B_c \to VV$ decays

| Decays | At $\omega$ = 0.40 GeV | | Using flavor dependent $\omega$ | |
|---|---|---|---|---|
| | $\Gamma_T / \Gamma$ | $\Gamma_L / \Gamma$ | $\Gamma_T / \Gamma$ | $\Gamma_L / \Gamma$ |
| $\Delta b = 0$, $\Delta C = -1$, $\Delta S = -1$ | | | | |
| $B_c^+ \to B^{*0} e \bar{\nu}_e$ | 0.21 | 0.79 | 0.30 | 0.70 |
| $B_c^+ \to B^{*0} \mu \bar{\nu}_\mu$ | 0.21 | 0.79 | 0.30 | 0.70 |
| $B_c^+ \to B_s^{*0} e \bar{\nu}_e$ | 0.14 | 0.86 | 0.20 | 0.80 |
| $B_c^+ \to B_s^{*0} \mu \bar{\nu}_\mu$ | 0.14 | 0.86 | 0.20 | 0.80 |
| $\Delta b = 1$, $\Delta C = 0$, $\Delta S = -1$ | | | | |
| $B_c^+ \to D^{*0} e \bar{\nu}_e$ | 0.44 | 0.56 | 0.47 | 0.53 |
| $B_c^+ \to D^{*0} \mu \bar{\nu}_\mu$ | 0.44 | 0.56 | 0.47 | 0.53 |
| $B_c^+ \to D^{*0} \tau \bar{\nu}_\tau$ | 0.41 | 0.59 | 0.50 | 0.50 |
| $\Delta b = 1$, $\Delta C = 1$, $\Delta S = 0$ | | | | |
| $B_c^+ \to J/\psi e \bar{\nu}_e$ | 0.10 | 0.90 | 0.12 | 0.88 |
| $B_c^+ \to J/\psi \mu \bar{\nu}_\mu$ | 0.10 | 0.90 | 0.12 | 0.88 |
| $B_c^+ \to J/\psi \tau \bar{\nu}_\tau$ | 0.10 | 0.90 | 0.11 | 0.89 |

## Table V. $|\psi(0)|^2$ and $\omega$ for vector and pseudoscalar mesons

| Meson | $|\psi(0)|^2$ (in GeV$^3$) | Parameter $\omega$ (in GeV) |
|---|---|---|
| $\rho(\pi)$ | 0.011 | 0.33 |
| $K^*(K)$ | 0.011 | 0.33 |
| $D^*(D)$ | 0.026 | 0.43 |
| $D_s^*(D_s)$ | 0.041 | 0.51 |
| $J/\psi(\eta_c)$ | 0.115 | 0.71 |
| $B^*(B)$ | 0.033 | 0.47 |
| $B_s^*(B_s)$ | 0.053 | 0.55 |
| $B_c$ | 0.281 | 0.96 |

**Tables VI. Form factors of $B_c \to V$ transition using flavor dependent $\omega$.**

| Modes | Transition | $V(0)$ | $A_0(0)$ | $A_1(0)$ | $A_2(0)$ |
|---|---|---|---|---|---|
| $\Delta b = 0, \Delta C = -1, \Delta S = -1$ | $B_c \to B_s^*$ | 5.19 | 0.57 | 0.79 | 3.24 |
| | $B_c \to B^*$ | 4.77 | 0.42 | 0.63 | 2.74 |
| $\Delta b = 1, \Delta C = 0, \Delta S = -1$ | $B_c \to D^*$ | 0.16 | 0.081 | 0.095 | 0.11 |
| | $B_c \to D_s^*$ | 0.29 | 0.16 | 0.18 | 0.20 |
| $\Delta b = 1, \Delta C = 1, \Delta S = 0$ | $B_c \to J/\psi(c\bar{c})$ | 0.91 | 0.58 | 0.63 | 0.74 |